\documentclass[10pt, letter, conference]{IEEEtran}
\IEEEoverridecommandlockouts
\usepackage{amsmath,amsfonts}
\usepackage{colortbl}
\usepackage{array} 
\usepackage{multirow}
\usepackage{multicol}
\usepackage{amsthm}
\usepackage{listings}
\usepackage[linesnumbered,ruled,vlined]{algorithm2e}
\usepackage{graphicx}
\usepackage{pgfplots}
\usepackage{textcomp}
\usepackage{xcolor}
\usepackage{cleveref}
\usepackage{makecell}
\usepackage{booktabs}
\usepackage{comment}
\usepackage{tabularx}
\usepackage{multirow}
\usepackage{bigdelim}
\usepackage{bm}
\usepackage{url}
\usepackage{xspace}
\usepackage{pifont}
\usepackage{arydshln}
\usepackage{verbatim}
\usepackage[referable]{threeparttablex}
\usepackage{calc} 
\usepackage{extarrows}
\usepackage{color}
\usepackage{float}
\usepackage{subcaption}
\usepackage{nicematrix}
\usepackage{geometry}
\usepackage{balance}
\usepackage[utf8]{inputenc}
\usepackage{hhline}
\usepackage{booktabs} 
\usepackage{tabularx} 
\usepackage{tikz}
\usetikzlibrary{shapes.geometric, arrows}
\tikzstyle{startstop} = [rectangle, rounded corners, minimum width=3cm, minimum height=1cm,text centered, draw=black]
\tikzstyle{io} = [trapezium, trapezium left angle=70, trapezium right angle=110, minimum width=3cm, minimum height=1cm, text centered, draw=black]
\tikzstyle{process} = [rectangle, minimum width=3cm, minimum height=1cm, text centered, draw=black]
\tikzstyle{arrow} = [thick,->,>=stealth]

\usepackage{filecontents}                                  
\usepackage{pgfplots}
\usepackage{pgfplotstable}
\usepackage{scalefnt}

\definecolor{USTgold}{RGB}{153,102,0}
\definecolor{USTyellow}{RGB}{204,153,0}
\definecolor{USTyellowlight}{RGB}{255,212,0}
\definecolor{USTorange}{RGB}{255,166,26}
\definecolor{USTpink}{RGB}{255,157,157}
\definecolor{USTblue}{RGB}{0,51,102}
\definecolor{USTmiddle}{RGB}{0,116,188}
\definecolor{USTlight}{RGB}{99,202,225}
\definecolor{USTgray}{RGB}{204,204,204}
\definecolor{USTred}{RGB}{237,27,47}
\definecolor{USTdarkred}{RGB}{124,35,72}

\definecolor{CUHKorange}{RGB}{244,106,18} 
\definecolor{CUHKblue}{RGB}{0,111,190}    
\definecolor{CUHKgreen}{RGB}{0,127,128}   
\definecolor{CUHKred}{RGB}{228,46,36}     
\definecolor{CUHKyellow}{RGB}{198,148,34} 
\definecolor{CUHKdark}{RGB}{114,44,114}   
\definecolor{CUHKmiddle}{RGB}{144,44,144} 
\definecolor{CUHKlight}{RGB}{167,44,167} 
\definecolor{lightblue}{RGB}{223, 235, 247}

\newcommand{\deftitle}[0]{{\texttt{E-morphic}}\xspace}
\newcommand{\deftextEsynTurbo}[0]{\mbox{\texttt{E-morphic}}\xspace}
\newcommand{\deftextEsyn}[0]{\mbox{\texttt{E-Syn}}\xspace}

\iftrue
\def\BibTeX{{\rm B\kern-.05em{\sc i\kern-.025em b}\kern-.08em
    T\kern-.1667em\lower.7ex\hbox{E}\kern-.125emX}}

\fi

\iftrue
\fi

\iftrue
\usepackage{titlesec}
\titlespacing\section{2pt}{5pt plus 1pt minus 1pt}{0pt plus 1pt minus 1pt}
\titlespacing\subsection{2pt}{5pt plus 1pt minus 1pt}{0pt plus 1pt minus 1pt}
\titlespacing\subsubsection{2pt}{5pt plus 1pt minus 1pt}{2pt plus 1pt minus 1pt}
\usepackage[inline]{enumitem}
\setlist{leftmargin=5.08mm}
\fi

\iftrue
\fi

\usepackage{cite}
\newcommand{\todo}[1]{\footnote{\textbf{\color{red}{TODO:}} #1}}

\def\BibTeX{{\rm B\kern-.05em{\sc i\kern-.025em b}\kern-.08em
    T\kern-.1667em\lower.7ex\hbox{E}\kern-.125emX}}
\newlist{myitemize}{itemize}{1}
\setlist[myitemize]{
  label=\textbullet, 
}

\usepackage{etoolbox}
\makeatletter
\patchcmd{\@makecaption}
  {\scshape}
  {}
  {}
  {}
\makeatother

\def\BibTeX{{\rm B\kern-.05em{\sc i\kern-.025em b}\kern-.08em
    T\kern-.1667em\lower.7ex\hbox{E}\kern-.125emX}}
\newcommand{\blackcircled}[1]{%
  \tikz[baseline=(char.base)]{%
    \node[shape=circle,draw,inner sep=1pt,fill=black] (char) {\textcolor{white}{\scriptsize #1}};%
  }%
}

\usepackage{amsmath,amsfonts}
\usepackage{colortbl}
\usepackage{array} 
\usepackage{multirow}
\usepackage{multicol}
\usepackage{amsthm}
\usepackage{listings}
\usepackage[linesnumbered,ruled,vlined]{algorithm2e}
\usepackage{graphicx}
\usepackage{pgfplots}
\usepackage{textcomp}
\usepackage{xcolor}
\usepackage{cleveref}
\usepackage{makecell}
\usepackage{booktabs}
\usepackage{comment}
\usepackage{tabularx}
\usepackage{multirow}
\usepackage{bigdelim}
\usepackage{bm}
\usepackage{url}
\usepackage{xspace}
\usepackage{pifont}
\usepackage{arydshln}
\usepackage{verbatim}
\usepackage[referable]{threeparttablex}
\usepackage{calc} 
\usepackage{extarrows}
\usepackage{color}
\usepackage{float}
\usepackage{subcaption}
\usepackage{nicematrix}
\usepackage{geometry}
\usepackage{balance}
\usepackage[utf8]{inputenc}
\usepackage{hhline}
\usepackage{booktabs} 
\usepackage{tabularx} 
\usepackage{tikz}
\usetikzlibrary{shapes.geometric, arrows}
\tikzstyle{startstop} = [rectangle, rounded corners, minimum width=3cm, minimum height=1cm,text centered, draw=black]
\tikzstyle{io} = [trapezium, trapezium left angle=70, trapezium right angle=110, minimum width=3cm, minimum height=1cm, text centered, draw=black]
\tikzstyle{process} = [rectangle, minimum width=3cm, minimum height=1cm, text centered, draw=black]
\tikzstyle{arrow} = [thick,->,>=stealth]

\usepackage{filecontents}                                  
\usepackage{pgfplots}
\usepackage{pgfplotstable}
\usepackage{scalefnt}

\begin{document}

\title{\deftitle: Scalable Equality Saturation  for Structural Exploration in Logic~Synthesis
}




%

\author{
    \IEEEauthorblockN{Chen Chen\textsuperscript{*}, Guangyu Hu\textsuperscript{* \dag}, Yuzhe Ma, Hongce Zhang\textsuperscript{\dag}}
    \IEEEauthorblockA{
        \textit{HKUST \& HKUST(GZ)} \\
        \{cchen099, yuzhema, hongcezh\}@connect.hkust-gz.edu.cn \\
        ghuae@connect.ust.hk
    }
    \and
    \IEEEauthorblockN{Cunxi Yu}
    \IEEEauthorblockA{
        \textit{University of Maryland, College Park} \\
        cunxiyu@umd.edu
    }
}


\maketitle
\begingroup
\renewcommand\thefootnote{}
\footnotetext{This work is supported in part by the National Natural Science Foundation of China under Grant No. 62304194, and by Guangzhou Municipal Science and Technology Project under Grant No. 2023A03J0013.}
\endgroup
\vspace*{-1cm}
\begingroup
\renewcommand\thefootnote{\textsuperscript{*}}
\footnotetext{Both authors contributed equally to this research.}
\endgroup
\begingroup
\renewcommand\thefootnote{}
\footnotetext{\dag 
 Corresponding author}
 \endgroup
\begin{abstract}
In technology mapping, the quality of the final implementation heavily relies on the circuit structure   after technology-independent optimization. 
Recent studies have introduced equality saturation as a novel optimization approach. 
However, its efficiency remains a hurdle against its wide adoption in logic synthesis.
This paper proposes a highly scalable and efficient framework named \deftextEsynTurbo. 
It is the first work that employs equality saturation for resynthesis after conventional technology-independent logic optimizations, enabling structure exploration before technology mapping. 
Powered by several key enhancements to the equality saturation framework, such as direct e-graph-circuit conversion, solution-space pruning, and simulated annealing for e-graph extraction, this approach not only improves the scalability and extraction efficiency of e-graph rewriting but also addresses the structural bias issue present in conventional logic synthesis flows through parallel structural exploration and resynthesis.
Experiments show that, compared to the state-of-the-art delay optimization flow in ABC, \deftextEsynTurbo on average achieves 12.54\% area saving   and  7.29\% delay reduction  on the large-scale circuits in the EPFL benchmark.

\end{abstract}

\section{Introduction}


Typically, technology mapping employs graph or tree covering algorithms to map an And-Inverter Graph (AIG) into standard cells.
However, technology mapping based on structural covering faces a challenge known as structural bias~\cite{structural_bias}. Namely, the quality of result (QoR) is largely dependent on the structure of subject graphs~\cite{lehman1997logic}.  If the initial structure is poor, the final mapping will also be sub-optimal, even with the use of heuristics or iterative recovery methods. 


To address the structural bias problem, previous research introduced lossless synthesis~\cite{techmap2006} to explore a few structural choices during mapping. 
It employs heuristic rewriting to  produce functional equivalent nodes that are then combined into a single subject graph with choices, which is subsequently used to derive the mapped netlist and researches~\cite{structural_bias}~\cite{DBLP:conf/date/StokSI99}~\cite{DBLP:journals/tcad/LehmanWGH97} targeted at mitigating structural bias heuristically, they cannot avoid this issue completely. 

\begin{figure}[t]
    \centering
    \includegraphics[width=0.49\textwidth]{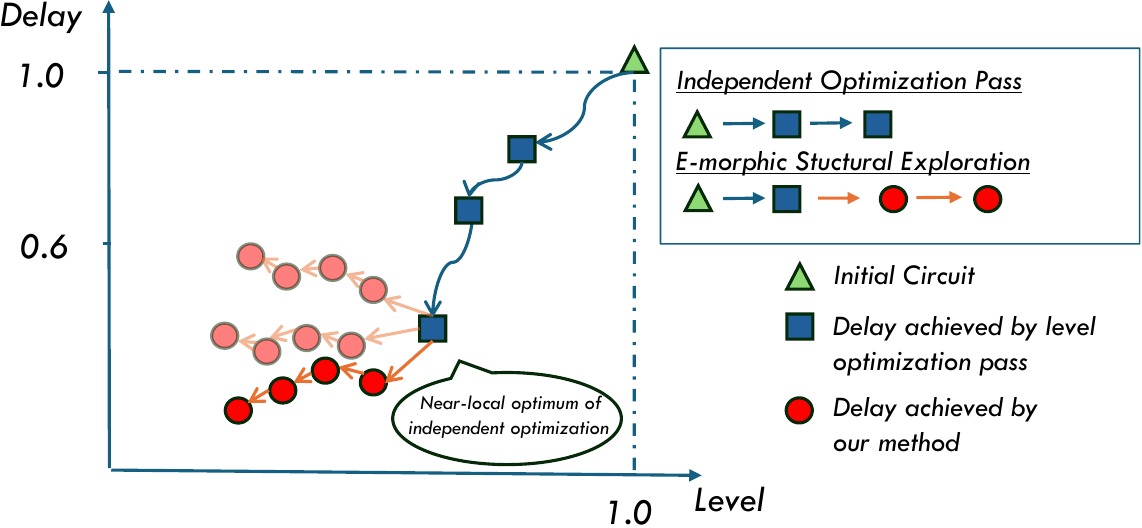}
    \caption{ \deftextEsynTurbo: parallel structural exploration for delay finetuning improves the circuit performance. }
    \label{fig:point_move}
    \vspace{-5pt}
\end{figure}

Recently, equality saturation has emerged as a promising technique, which uses a non-destructive rewriting approach to achieve better Pareto-optimal design space exploration and has been widely applied in the EDA field~\cite{ustun2023equality}.  The applications include program optimization~\cite{EQsaturation}, compiling optimization~\cite{cheng2023seer,ESFO}, datapath optimization~\cite{coward2023automating}, multiplier optimization~\cite{Wanna2023MultiplierOV}, etc.
Among these prior works, \deftextEsyn~\cite{chen2024syn} first applied e-graph rewriting to Boolean logic optimization and demonstrated the potential of e-graphs in exploring optimal logic structure. 
However, due to the complexity of e-graph-based optimization, these previous works have to be on a limited scale, with no more than 40,000 e-nodes~\cite{coward2023automating}.
When the graph scale significantly grows, the two main steps in equality saturation, which are rewriting (to expand solution space) and extraction (to pick a good solution), would inevitably suffer from excessively long runtime.
Therefore, it is still an unanswered question whether e-graphs can play an effective role in large-scale Boolean logic optimization scenarios.
As we work to scale up equality saturation for logic synthesis, we acquire the following two key insights.
$(1)$ The method ~\cite{structural_bias}~\cite{techmap2006} faces a primary challenge: because \texttt{ABC} inherently performs local rewriting~\cite{DBLP:conf/dac/MishchenkoCB06}, previous approaches do not maintain equivalence classes during the rewriting process. Instead, they create equivalence classes by detecting equivalent nodes through simulation and SAT checking across multiple circuits after performing local rewrites against the original circuit. This approach maintains only one additional equivalent node per equivalence class in the most commonly used \texttt{ABC} operator \texttt{dch} and has a restricted ability to make global structural changes. However, e-graphs utilize congruence closure-based~\cite{bradley2007calculus} non-destructive rewriting, fundamentally aiming to maintain a vast number of equivalence classes. \textbf{By performing only a small number of iterations of e-graph-based rewriting, e-graphs can generate significantly more equivalence classes than previous methods, enabling larger structural transformations. Additionally, using fewer iterations ensures that independent optimization targets such as node sizes and logic levels are not adversely affected.}  Through a case study presented in Figure~\ref{fig:point_move}, we observed that as technology-independent optimizations approach a near-local optimum, our tool, \deftextEsynTurbo, employs parallel e-graph-based resynthesis to mitigate structural bias issues. This approach facilitates the exploration of more optimal solutions for technology mapping targets.
$(2)$ Although fewer iterations already generate an excessive number of equivalence classes and nodes, the presence of too many equivalent nodes within each class severely impacts the time required for e-graph extraction. \textbf{Therefore, it is important to have an efficient and effective e-graph extraction method to explore various structural choices within a reasonable time.  
To this end, we incorporate a series of new features into the equality saturation framework, including solution space pruning and simulated annealing. Solution space pruning effectively reduces the over-abundance of redundant equivalent nodes in each equivalence class, significantly decreasing the time required for e-graph extraction while simulated annealing enables the escape from local minima when selecting from equivalent terms.} 
Bearing these ideas in mind, we build an equality saturation framework named \deftextEsynTurbo for practical Boolean logic synthesis problems. It uses equality saturation in a fast and scalable manner to address the efficiency and scalability issues. The word ``morphic'' in its name suggests its intended role in logic synthesis --- exploring diverse logic structures to improve the post-mapping design quality.

The contributions of this paper are as follows:
\begin{itemize}
    \item To our best knowledge, this is the first work that adopts e-graph rewriting to address the structural bias issue of technology mapping in logic synthesis. 
    \item It brings up several important updates to the equality  saturation framework, such as DAG-to-DAG conversion, solution space pruning, simulated annealing, and multi-batch parallel computing for e-graph extraction. These techniques notable enhance the efficiency of equality  saturation  in logic synthesis and could be helpful for other e-graph applications.
    \item We implement a scalable and effective logic optimization framework and achieve 12.54\% area saving  and 7.29\% delay reduction compared to a competitive delay-optimal synthesis flow on the large-scale circuits in the EPFL benchmark.\looseness=-1
\end{itemize}

\section{Background}

\subsection{E-Graph and Equality Saturation}
\begin{figure}[htbp]
  \centering
  \includegraphics[width=0.3\textwidth]{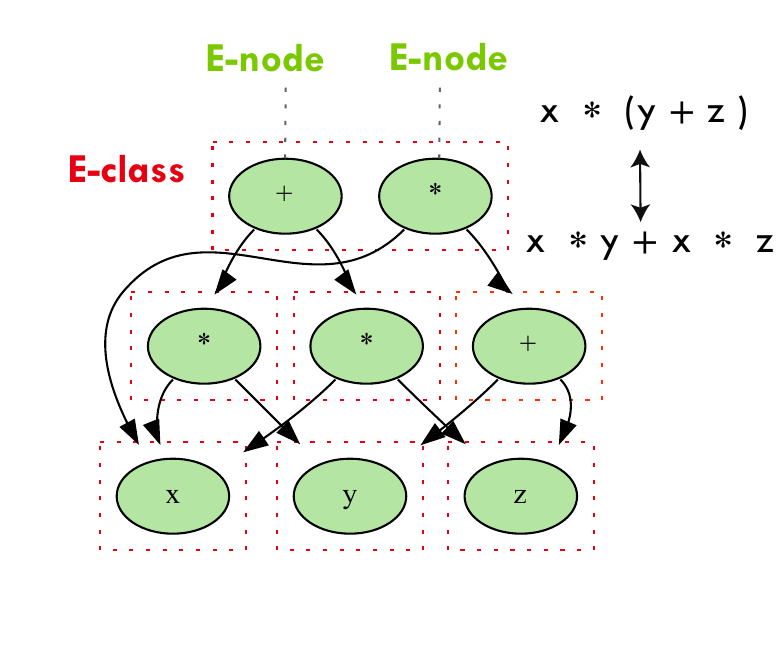}
    \vspace{-2em}
  \caption{This e-graph represents two equivalent expressions. Each green node is an e-node, and each red-dotted box is an e-class. Edges connect e-nodes to child e-classes.}
  \label{fig:egraph-example}
\end{figure}

An e(quivalence)-graph is a data structure used to represent a set of equivalent terms (expressions).
\Cref{fig:egraph-example} gives an example of an e-graph where each green node is called an e-node that represents a term from a given language. The red-dotted boxes are equivalent classes (e-classes) where e-nodes in the same class represent equivalent expressions. E-graph also maintains congruence relations \footnote{If two terms $a$ and $b$ are equivalent, any function applications over $a$ and $b$ would also be equivalent ($f(a) \equiv f(b)$).}


For a given input term, we may rewrite it into other equivalent forms following a set of rules (such as those listed in \Cref{tab:rewrites}). This process of finding equivalent terms is called equality saturation.
Given a term $\mathcal{T}$, an initial e-graph $\mathcal{E}$ is constructed. Then we can iteratively find patterns that match with the left-hand-side of rewriting rules and record the right-hand-side patterns that will be added to the graph. 
Specifically, the rewriting does not remove the original term but only adds information to $\mathcal{E}$. Therefore, it is non-destructive and the ordering of rewriting will not affect the outcome. This solves the so-called phase ordering problem in optimization.
The rewriting iterations may terminate when no more rule is applicable (namely, the graph is saturated) or when a certain exiting condition is met. Then an optimized term can be extracted from the final e-graph where the optimality of terms is measured by a cost function defined w.r.t. the whole graph. \looseness=-1

\begin{table}
  \centering
  \caption{Examples of rewriting rules}
  \label{tab:rewrites}
  \resizebox{0.85\columnwidth}{!}{
  \begin{tabular}{c|c|c}
    \hline
    \textbf{Class} & \textbf{Pattern (LHS)} & \textbf{Transformation (RHS)} \\
    \hline
    \multirow{2}{*}{\textbf{Commutativity}} & $a * b$ & $b * a$ \\
    & \cellcolor{lightblue}$a + b$ & \cellcolor{lightblue}$b + a$ \\
    \hline
    \multirow{2}{*}{\textbf{Associativity}} & $(a * b) * c$ & $a * (b * c)$ \\
    & \cellcolor{lightblue}$(a + b) + c$ & \cellcolor{lightblue}$a + (b + c)$ \\
    \hline
    \multirow{3}{*}{\textbf{Distributivity}} & $a * (b + c)$ & $a * b + a * c$ \\
    & \cellcolor{lightblue}$(a + b) * (a + c)$ & \cellcolor{lightblue}$a + (b * c)$ \\
    & $(a * b) + (a * c)$ & $a * (b + c)$ \\
    \hline
    \multirow{2}{*}{\textbf{Consensus}} & $(a * b) + ((\neg a) * c) + b * c$ & $(a * b) + (\neg a) * c$ \\
    & \cellcolor{lightblue}$((a + b) * ((\neg a) + c)) * (b + c)$ & \cellcolor{lightblue}$(a + b) * ((\neg a) + c)$ \\
    \hline
    \multirow{2}{*}{\textbf{De-Morgan}} & $\neg(a *b )$ & $\neg a + \neg b$ \\
    & \cellcolor{lightblue}$\neg(a + b)$ & \cellcolor{lightblue}$(\neg a) * (\neg b)$ \\
    \hline
  \end{tabular}
  }
\end{table}



\begin{figure}[htbp]
\centering
\includegraphics[width=0.49\textwidth]{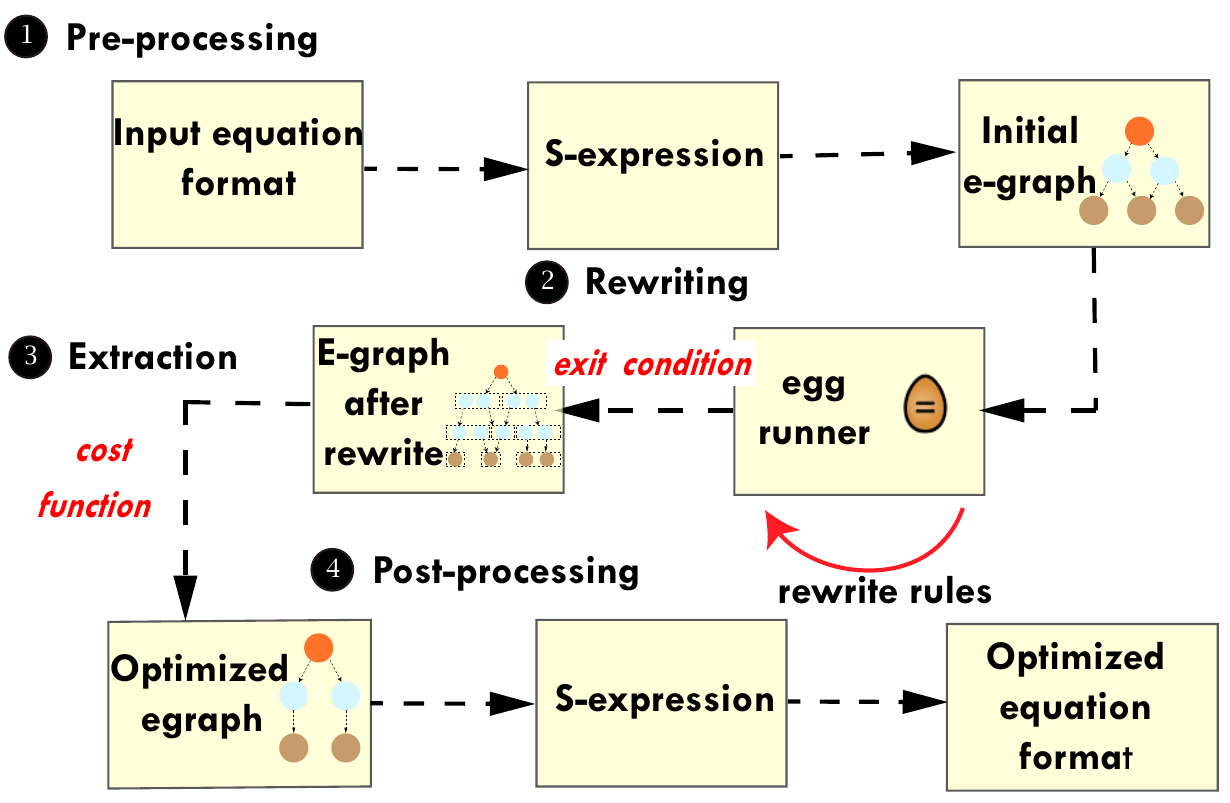}
\caption{The workflow of \deftextEsyn~\cite{chen2024syn}}
\label{fig:E-syn}
\end{figure}

\subsection{E-Graph Rewriting in Logic Synthesis}

E-graph has demonstrated its potential in many optimization problems~\cite{cheng2023seer,ESFO,coward2023automating,Wanna2023MultiplierOV,ustun2022impress}, while the prior work \deftextEsyn~\cite{chen2024syn} first applied equality saturation in logic synthesis. \deftextEsyn is based on an existing e-graph processing framework \texttt{egg}~\cite{egg}, as presented by \Cref{fig:E-syn}.
Its workflow is as follows: \blackcircled{1} \textbf{Pre-processing:} \deftextEsyn takes advantage of the ABC logic synthesis tool to transform input Verilog or AIG into the equation format. Then \deftextEsyn converts the equation format into the S-expression, which is a data structure for nested lists commonly used in Lisp-like programming languages and can be converted into e-graphs using the built-in functionality of \texttt{egg}.
\blackcircled{2} \textbf{Rewriting:} The \texttt{egg} runner will iteratively apply a set of rewriting rules until the e-graph is saturated or an exit condition is met, such as when the time limit or maximum iteration is reached. \blackcircled{3} \textbf{Extraction:} 
It selects the best term based on a cost function. To achieve overall algorithmic efficiency, a commonly used approach is to traverse the e-graph bottom-up and greedily select the e-node with the lowest cost among the equivalent ones in the same e-class.
\blackcircled{4} \textbf{Post-processing:} \deftextEsyn finally converts the optimized e-graph back into the equation format, which can be mapped to gate-level circuits using synthesis tools. Then the performance metrics of the circuit will be reported.

\section{Methodology}\label{sec:our-method}

\deftextEsynTurbo effectively addresses several significant challenges associated with previous works~\cite{chen2024syn,coward2022automatic}, which struggles with problems like inefficient intermediate representations and restricted cost functions, leading to scalability issues. Our solution overcomes these limitations through critical advancements, which we will discuss in the following subsections.

\subsection{\deftextEsynTurbo Overview}

Our framework, \deftextEsynTurbo, introduces a novel and efficient method for applying e-graphs in logic synthesis. We identify a unique role for e-graph-based optimization, between technology-independent optimization and technology mapping. \deftextEsynTurbo offers several key advantages:

\begin{enumerate}
    \item \textbf{Structural Exploration before Technology Mapping:} Parallel E-graph-based resynthesis using fewer rewriting iterations after technology-independent optimization enables efficient structural exploration to address the structural bias issue.
    
    \item \textbf{Novel Extraction Method:} Integrating Solution Space Pruning and Simulated Annealing-Based Algorithms with Multi-Threaded Parallel Computing for Efficient E-graph Extraction and Faster Evaluation.
    
    \item \textbf{Efficient Implementation:} Our framework incorporates several efficiency-enhancing features: (a) direct DAG-to-DAG conversion between circuits and e-graphs, eliminating the overhead of intermediate S-expressions, and (b) machine learning-based cost estimation for rapid quality assessment.
\end{enumerate}

\begin{figure*}[!thb]
    \centering
    \includegraphics[width=0.8\linewidth]{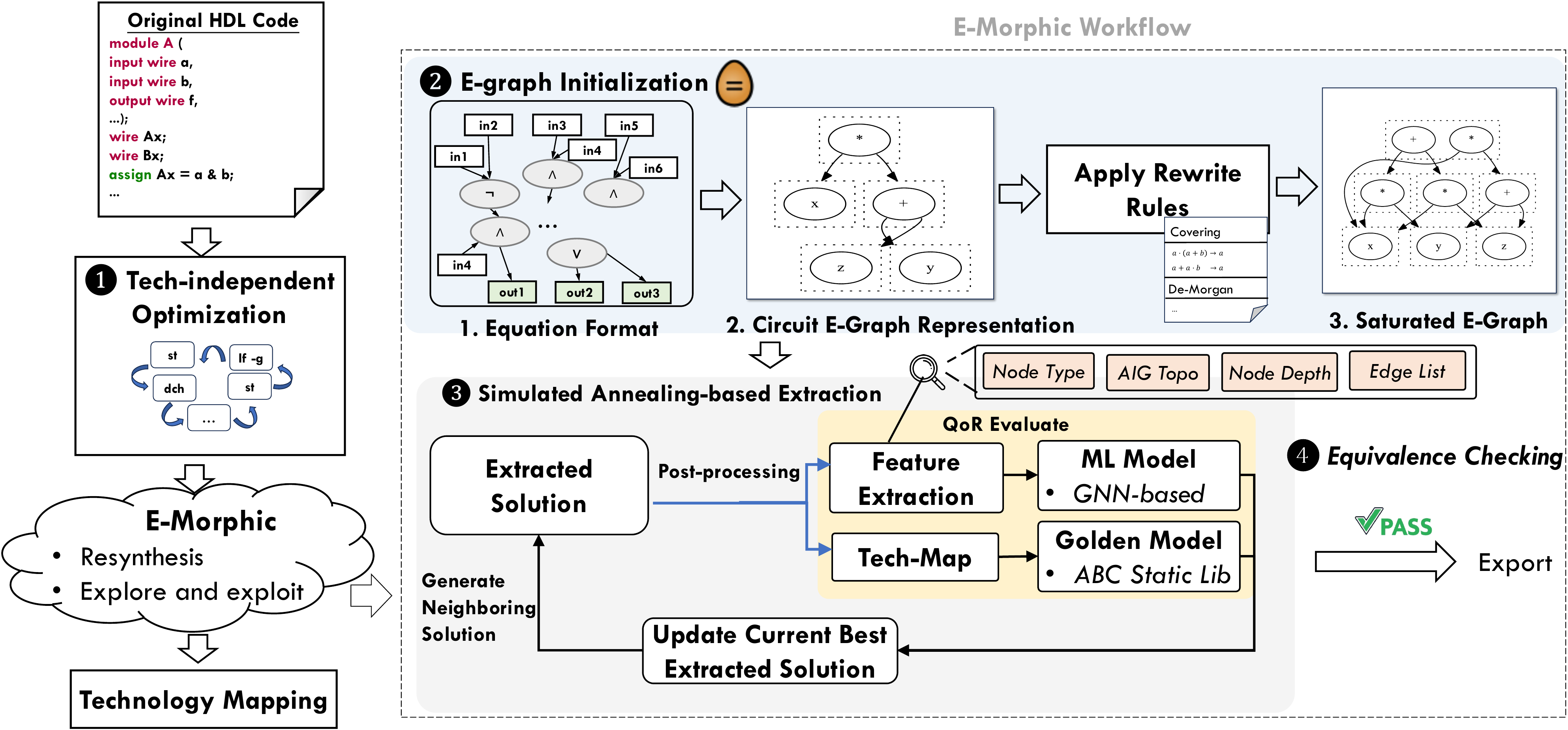}
    \caption{\deftextEsynTurbo overview.}
    \label{EsynTurbo_overview}
    \vspace{-10pt}
\end{figure*}

Figure~\ref{EsynTurbo_overview} illustrates the workflow of \deftextEsynTurbo. The input circuit first undergoes conventional technology-independent optimization (\blackcircled{1}). Our efficient DAG-to-DAG conversion (\blackcircled{2}) then initializes the e-graph. The highlight of our approach lies in the extraction and evaluation phase (\blackcircled{3}), with our novel simulated annealing algorithm, parallel processing, and dual cost evaluation methods (ABC-based quality-prioritized  and GNN-based rapid estimation). 


\begin{figure}[htbp]
    \centering
    \includegraphics[width=0.49\textwidth]{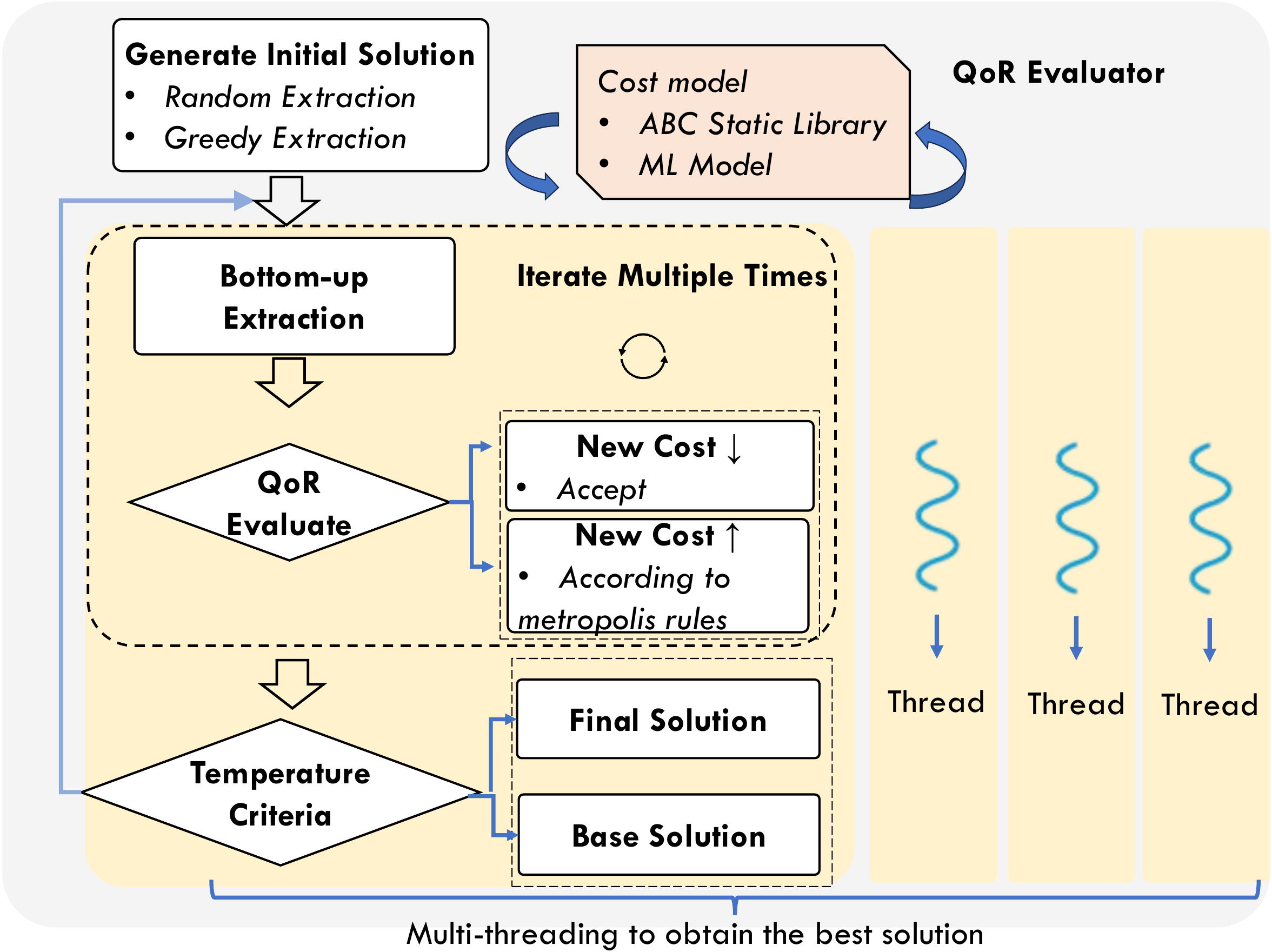}
    \caption{Simulated-annealing-based extraction algorithm.}
    \label{fig:SA}
\end{figure}

\subsection{Efficient Extraction methods}
\label{sec:faster_extraction}
\subsubsection{Simulated Annealing e-graph Extraction}
The extraction phase in e-graph-based optimization is crucial for obtaining high-quality results. However, extraction from equality saturated graphs is known to be NP-hard~\cite{Baader_Nipkow_1998}, and traditional greedy algorithms often fail to reflect QoR after circuit technology mapping. To address this challenge, we have developed a novel extraction algorithm based on simulated annealing (SA)~\cite{kirkpatrick1983optimization}, a probabilistic technique for approximating the global optimum of a given function. In addition, we integrate Solution Space Pruning and Multi-Threaded Parallel Computing to achieve efficient e-graph extraction and facilitate broad structural exploration.

Our SA extractor combines the principles of bottom-up extraction with simulated annealing to efficiently explore the solution space of an e-graph. This hybrid approach aims to overcome local optima and find high-quality circuit representations. The algorithm (illustrated in~\Cref{fig:SA}) operates as follows:

We begin by generating an initial solution using a basic extraction method, such as greedy depth-minimization or random selection. The main loop of the algorithm then iteratively improves this solution while gradually decreasing the temperature parameter of simulated annealing, with specific cooling schedules detailed in \Cref{sec:exp_setting}. At each iteration, we generate a neighboring solution, evaluate its cost, and decide whether to accept or reject it based on the cost difference and current temperature. This process allows for occasional uphill moves, enabling the algorithm to escape local optima.
\vspace{-3pt}

The neighbor generation process, illustrated in~\Cref{alg:generate_neighbor}, starts from leaf nodes and propagates changes upwards through the e-graph. For each node, we decide whether to update its selection based on a combination of cost improvement and random chance. This randomness allows for exploration of different parts of the solution space while ensuring that the solution quality does not degrade below the initial solution.

\begin{algorithm}[!t]
  \caption{Generate Neighboring Solution}
  \label{alg:generate_neighbor}
  \SetAlgoLined
  \KwIn{Current solution $\mathcal{O}_{current}$, e-graph $\mathcal{E}$, cost function $\mathcal{C}$, random probability $p_{random}$}
  \KwOut{New solution $\mathcal{O}_{new}$}
  
  $\mathcal{O}_{new} \leftarrow \mathcal{O}_{current}$\;
  $traversalQueue \leftarrow$ LeafNodes($\mathcal{E}$)\;
  $Costs\_map \leftarrow$ \{ $\infty$ , for all e-classes\}\;
  \While{$traversalQueue$ is not empty}{
    $enode \leftarrow traversalQueue$.pop()\;
    $class \leftarrow$ GetClass($enode$)\;
    $prev\_cost \leftarrow Costs\_map[class]$\; 
    $CS \leftarrow \text{all children of } enode$\;
    \If{$\mathcal{C}$ is a ``sum cost''}{
        $new\_cost \leftarrow \mathcal{C}(enode) + \sum_{i \in CS} Costs\_map[i]$\; 
    }
    \If{$\mathcal{C}$ is a ``depth cost''}{
        $new\_cost \leftarrow \mathcal{C}(enode) + \max_{i \in CS} \{Costs\_map[i]\}$\;
    }
    \If{$prev\_cost = \infty$ \textbf{or} ($new\_cost < prev\_cost$ \textbf{and} $\text{random}() \geq p_{random}$)}{
      $\mathcal{O}_{new}$.choose($class$, $enode$)\; 
      $Costs\_map[class] \leftarrow new\_cost$\; 
      $traversalQueue$.extend(GetParents($class$))\;
    }
  }
  \Return $\mathcal{O}_{new}$\;
\end{algorithm}

\subsubsection{Solution Space Pruning for Extraction}
\begin{figure}[thb]
    \centering
    \includegraphics[width=0.4\textwidth]{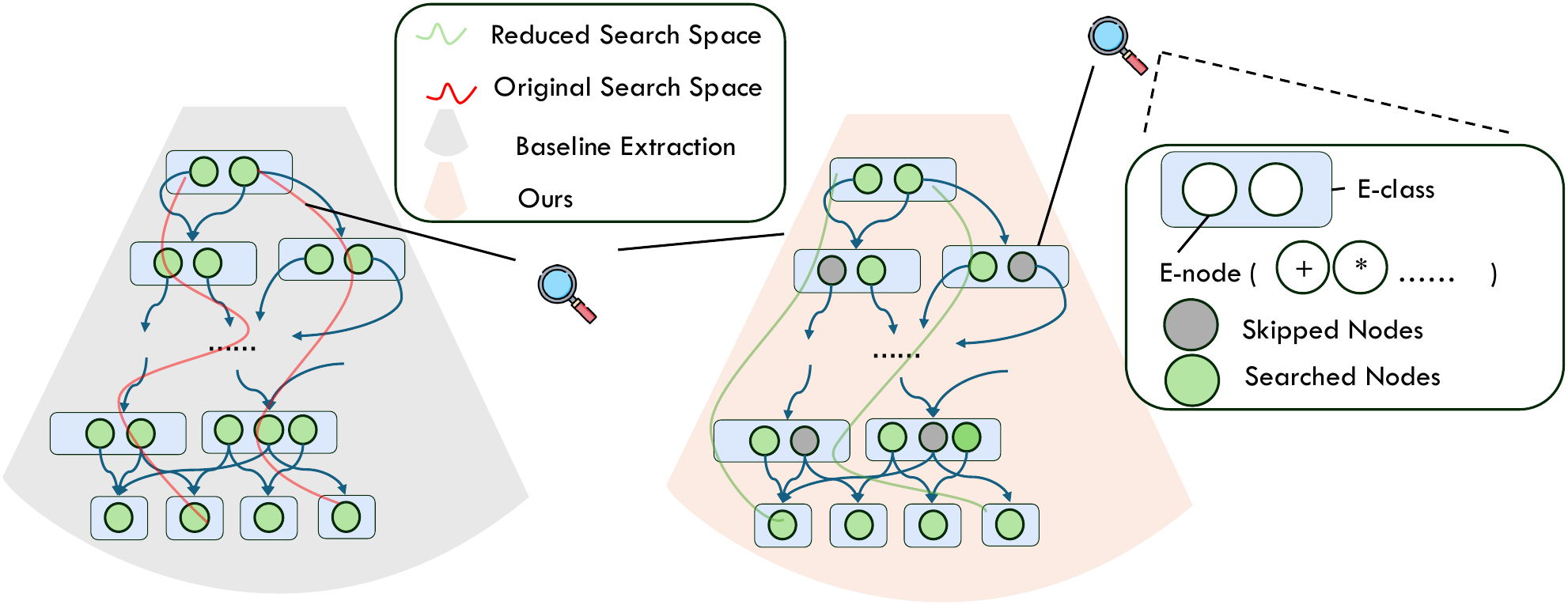}
    \caption{Solution space pruning.}
    \label{fig:slv_space_comp}
\end{figure}

Additionally, we employ solution space pruning techniques to accelerate the extraction process.
Traditional bottom-up extraction algorithms often traverse every e-node in the e-graph, recalculating costs even when changes are unlikely to yield improvements. 
In contrast, our method maintains a traversal queue in~\Cref{alg:generate_neighbor} that only records the e-nodes with costs less than or equal to the minimum cost in each e-class, effectively skipping the remaining e-nodes with higher costs, thereby significantly filtering out the numerous redundant e-nodes introduced in each e-class due to commutative and associative laws as shown in \Cref{fig:slv_space_comp}. In addition, the minimum cost of each e-class is maintained by a hashmap called $Costs\_map$, which caches the optimal costs per e-class. This approach avoids redundant computations by preventing the re-evaluation of e-node costs when the costs of their child nodes remain unchanged.

By combining these techniques, we achieve a more efficient and effective extraction process that is particularly well suited for large and complex e-graphs encountered in large circuit optimization scenarios.

\subsubsection{Multithreaded Parallel Cost Evaluation}

In the QOR evaluation phase of our simulated annealing algorithm as referenced in~\Cref{fig:SA}, we support multi-threading to enhance efficiency. Leveraging initial solutions derived from bottom-up extraction, we concurrently run multiple annealing processes across different threads. This parallel execution allows us to perform QOR fine-tuning on multiple solution sets simultaneously, facilitating extensive resynthesis for a variety of structural transformations. Ultimately, we map all the parallel-generated solutions and select the one with the best QOR results.\looseness=-1
\subsection{Dual-Model Approach for Cost Estimation}
\label{sec:cost_models}

To balance optimization speed and result quality, we employ two complementary cost models in our extraction process:

\begin{enumerate}
    \item \textbf{Runtime-prioritized mode:} This mode primarily uses a Graph Neural Network (GNN) model for rapid estimation. The GNN quickly evaluates circuit characteristics based on node type, topology order, and connection relationships. While less accurate, it provides fast feedback to guide the simulated annealing process, making it particularly useful for exploring many candidate solutions.
    
    \item \textbf{Quality-prioritized mode:} This mode performs a fast but rough mapping to obtain a QoR estimation. It converts the extracted circuit to an equation file, processes it using \texttt{ABC} with a standard cell library, and applies technology mapping and optimizations. The resulting circuit delay serves as the primary cost metric, ensuring high-quality evaluations at the cost of longer runtime.
\end{enumerate}

By combining these modes, our approach efficiently navigates the complex solution space of e-graphs, balancing speed and accuracy in cost estimation for various circuit optimization tasks.

\subsection{Efficient Implementation}
\label{sec:efficient_impl}

Building upon our novel approach for structural exploration and extraction, we have developed several key techniques to enhance the efficiency of our framework. These improvements address the major bottlenecks in previous e-graph-based logic synthesis methods, allowing \deftextEsynTurbo to handle larger and more complex circuits.

\subsubsection{Direct DAG-to-DAG Conversion}
\label{sec:conversion}

\begin{figure}[thb]
    \centering
    \frame{\includegraphics[width=0.99\linewidth]{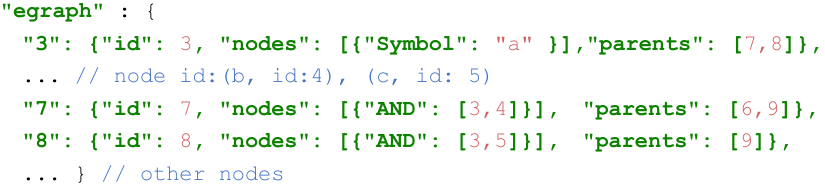}}
    \caption{The intermediate format for efficient e-graph---circuit conversion in pre- and post-processing.}
    \label{fig:circuit2egraph}
\end{figure}
\begin{figure}[thb]
    \centering
    \includegraphics[width=0.49\textwidth]{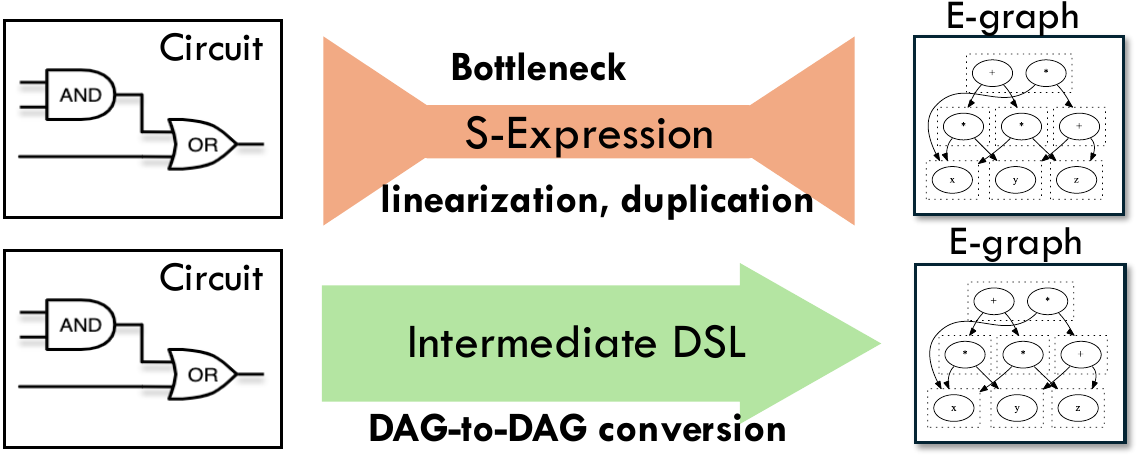}
    \caption{Prior and our new e-graph--circuit conversion.}
    \label{fig:conversion}
\end{figure}

Previous work in e-graph-based logic synthesis~\cite{chen2024syn} relied on S-expressions as the intermediate representation between circuits and e-graphs. However, S-expressions, being nested lists derived from flattening abstract syntax trees (ASTs), are not ideal for representing circuit graphs. Shared nodes must be duplicated in S-expressions, leading to significant space and time overhead during circuit-to-e-graph conversion and potentially degrading the overall quality of results (QoR). To address this limitation, we have developed a more efficient intermediate format that serves as a bridge for conversion.  It eliminates the need for complex parsing and reconstruction steps, significantly reducing computational overhead as illustrated in~\Cref{fig:conversion}. This format is essentially a serialization of the initial e-graph, using unique identifiers to represent nodes and their relationships. Figure \ref{fig:circuit2egraph} provides an example of this representation. For an equation format input, each assignment corresponds to a node in the initial e-graph, which is referred to by an \texttt{id} in our DSL. The one-to-one correspondence between circuit elements and their e-graph counterparts avoids the exponential growth in representation size of S-expressions.

\subsubsection{Flexible Cost Model Integration}
Our framework provides an interface for integrating various cost models, including machine learning-based approaches. We don't limit on the specifics of the particular model, this flexibility allows us to leverage state-of-the-art techniques for rapid quality assessment. Later,  Section~\ref{sec:gnn-based-cost-model} will demonstrate the use of a GNN model to provide rapid cost estimation to guide the simulated annealing process.

By combining these efficiency-enhancing techniques, \deftextEsynTurbo achieves significant speed-ups, enabling the application of e-graph-based optimization to a wider range of circuit sizes and complexities.




\begin{table*}
    \caption{QoR and runtime comparison between \deftextEsynTurbo and the baseline.}
    \centering
    \resizebox{\linewidth}{!}{%
    \begin{tabular}{l|rrrr|rrrr|rrrc}
    \hline
    \multicolumn{1}{c|}{\multirow{2}{*}{Circuit}} &
    \multicolumn{4}{c|}{\textbf{\textit{SOP Balancing Baseline}} } & 
    \multicolumn{4}{c|}{\textbf{\textit{SOP Balancing  + \deftextEsynTurbo w/o ML model}}} & 
    \multicolumn{4}{c}{\textbf{\textit{SOP Balancing  + \deftextEsynTurbo w ML model}}}
    
    \\ \cline{2-13} 
     &
    \multicolumn{1}{c}{Area ($\mu m^2$)} & 
    \multicolumn{1}{c}{Delay ($ps$)} & 
    \multicolumn{1}{c}{lev } & 
    \multicolumn{1}{c|}{runtime ($s$)} & 
    \multicolumn{1}{c}{Area ($\mu m^2$)} & 
    \multicolumn{1}{c}{Delay ($ps$)} & 
    \multicolumn{1}{c}{lev } & 
    \multicolumn{1}{c|}{runtime ($s$)} & 
    \multicolumn{1}{c}{Area ($\mu m^2$)} & 
    \multicolumn{1}{c}{Delay ($ps$)} & 
    \multicolumn{1}{c}{lev } & 
    \multicolumn{1}{c}{runtime ($s$)}
 \\
    \hline
      hyp & 370305.16 & 253779.06 & 10929 & 1496.23 & 386483.84 & \cellcolor{lightblue}\textbf{240435.12} & 10948 & 2102.52 & \cellcolor{lightblue}\textbf{366751.38} & \cellcolor{lightblue}\textbf{248781.94} & 11033 & 1646.86  \\ \hline
    div & 56873.2 & 25378.59 & 1258 & 108.04 & \cellcolor{lightblue}\textbf{34475.98} & \cellcolor{lightblue}\textbf{24747.77} & 1300 & 280.79 & \cellcolor{lightblue}\textbf{48386.24} & \cellcolor{lightblue}\textbf{24333.65} & 1298 & 147.77  \\ \hline   
    mem\_ctrl & 29433.64 & 1281.99 & 61 & 76.5 & \cellcolor{lightblue}\textbf{29219.49} & \cellcolor{lightblue}\textbf{1056.09} & \cellcolor{lightblue}\textbf{59} & 113.17 & \cellcolor{lightblue}\textbf{29365.29} & \cellcolor{lightblue}\textbf{1081.48} & \cellcolor{lightblue}\textbf{60} & 97.09  \\ \hline        
    log2 & 41154.32 & 5233.86 & 231 & 161.49 & \cellcolor{lightblue}\textbf{32840.69} & \cellcolor{lightblue}\textbf{5033.46} & 240 & 289.79 & 41234.11 & \cellcolor{lightblue}\textbf{5153.74} & 233 & 208.30   \\ \hline
multiplier & 35799.38 & 2632.08 & 147 & 54.68 & \cellcolor{lightblue}\textbf{31443.34} & \cellcolor{lightblue}\textbf{2371.24} & \cellcolor{lightblue}\textbf{130} & 105.02 & \cellcolor{lightblue}\textbf{35002.03} & \cellcolor{lightblue}\textbf{2599.37} & 147 & 87.15  \\ \hline
        sqrt & 53262.02 & 78627.99 & 3574 & 112.27 & \cellcolor{lightblue}\textbf{33852.66} & \cellcolor{lightblue}\textbf{72796.83} & 3578 & 270.21 & \cellcolor{lightblue}\textbf{48909.25} & \cellcolor{lightblue}\textbf{76432.02} & 3579 & 143.43  \\ \hline
    square & 21240.14 & 834.01 & 69 & 25.12 & 21418.84 & \cellcolor{lightblue}\textbf{792.09} & 69 & 50.22 & 21532.44 & \cellcolor{lightblue}\textbf{794.16} & 70 & 37.46  \\ \hline
    arbiter & 4464.75 & 288.18 & 23 & 12.58 & 4587.45 & \cellcolor{lightblue}\textbf{270.98} & \cellcolor{lightblue}\textbf{22} & 23.34 & 4532.63 & \cellcolor{lightblue}\textbf{284.27} & 24 & 20.18  \\ \hline
    sin & 16052.7 & 3712.41 & 119 & 34.95 & \cellcolor{lightblue}\textbf{15737.77} & \cellcolor{lightblue}\textbf{3567.12} & \cellcolor{lightblue}\textbf{113} & 62.31 & \cellcolor{lightblue}\textbf{15737.77} & \cellcolor{lightblue}\textbf{3685.34} & \cellcolor{lightblue}\textbf{118} & 50.83  \\ \hline
    adder & 1206.99 & 584.53 & 57 & 1.96 & 1282.11 & \cellcolor{lightblue}\textbf{536.42} & 57 & 10.2 & 1285.61 & \cellcolor{lightblue}\textbf{546.86} & 57 & 7.07  \\ \hline
        \textcolor{blue}{GEOMEAN} & 25274.02 & 5620.01 & 292 & 60.32 & \cellcolor{lightblue}\textbf{22104.32} & \cellcolor{lightblue}\textbf{5210.55} & \cellcolor{lightblue}\textbf{287} & 126.25 & \cellcolor{lightblue}\textbf{24660.84} & \cellcolor{lightblue}\textbf{5390.13} & 295 & 91.42  \\ \hline
\textcolor{red}{Improvements} &             &             &             &  &\textbf{12.54\%}             &    \textbf{7.29\%}           &  &            & \textbf{2.42\%} & \textbf{4.09\%}  & &  \\ \hline
\end{tabular}}
\label{tab:all-results}
\vspace{-10pt}
\end{table*}

\section{Experiments}\label{sec:experiment}
\subsection{Experiment Setting}\label{sec:exp_setting}
We implement the \deftextEsynTurbo framework in \texttt{Rust} to interface with the existing \texttt{egg} library for e-graphs. 
All experiments are performed on a Ubuntu 20.04.4 LTS server with dual Intel Xeon Platinum 8375C processors and 256GB memory.
The test circuits all come from the EPFL benchmark suite~\cite{amaru2015epfl},
containing 2 small-, 2 medium- and 6 large-scale cases.
Throughout these experiments, we use ASAP $7 nm$~\cite{clark2016asap7} technology library to evaluate the post-mapping QoR.
E-graph rewriting takes 5 iterations, which already produces a substantial number of equivalence classes to facilitate structural exploration.
In the quality-prioritized mode, we utilize 4 threads, with the annealing exit condition set to 4 iterations.
In the first iteration, the temperature \( T \) is set to 2000 as a high temperature \( T_1 \)  encourages acceptance of inferior solutions, helping to avoid local optima. For the 2nd and 3rd iterations, the temperature is defined as \( T_n = T_{n-1} \times \frac{|\text{new cost} - \text{old cost}|}{n \times 10000} \) (\( n \) is the number of iterations), ensuring progression into the pseudogreedy local-search stage. In the 4th iteration, the temperature is adjusted to \( T_n  = T_{n-1} \times \frac{|\text{new cost} - \text{old cost}|}{n} \), allowing for further enhancements in solution quality. For the runtime-prioritized mode, we utilize 6 threads to balance the performance loss of the cost model. All the circuits optimized through \deftextEsynTurbo are verified for equivalence using the \texttt{cec} command in \texttt{ABC}.

\begin{table}[!t]
\centering
\caption{Comparison of  e-graph--circuit conversion with a time-limit of 3600 seconds and 8 GB memory-limit. ``Forward'' means circuit to e-graph conversion and ``backward'' refers to e-graph to circuit conversion. \textcolor{red}{TO} represents ``timeout.'' \textcolor{red}{MO} stands for ``out-of-memory.''}
\label{tab:time-data} 
\resizebox{\linewidth}{!}{
\begin{tabular}{|c|c|c|c|c|c|}
\hline
\multicolumn{2}{|c|}{\textbf{Design}} & \multicolumn{2}{c|}{\textbf{\deftextEsyn\cite{chen2024syn}}} & \multicolumn{2}{c|}{\textbf{\deftextEsynTurbo (this work)}} \\ \cline{1-6} 
\textbf{Name} & \textbf{\#. e-node} & \textbf{Forward ($s$)} & \textbf{Backward ($s$)} & \textbf{Forward ($s$)} & \textbf{Backward ($s$)}  \\ \hline
hyp         &   420897    & \textcolor{red}{TO \& MO}              & \textcolor{red}{N.A.}$^{\star}$& \textbf{8.2}                &  \textbf{4.6}               \\ \hline
div         &   101860    & \textcolor{red}{TO \& MO}              & \textcolor{red}{N.A.}$^{\star}$& \textbf{1.57}          & \textbf{1.2}          \\ \hline
mem\_ctrl    &  84701     & \textcolor{red}{TO \& MO}              & \textcolor{red}{N.A.}$^{\star}$& \textbf{1.44}          & \textbf{0.8}          \\ \hline
log2       &    54532    & \textcolor{red}{TO \& MO}              & \textcolor{red}{N.A.}$^{\star}$& \textbf{0.93}           & \textbf{0.71}          \\ \hline

multiplier &   50761     & \textcolor{red}{TO}              & \textcolor{red}{N.A.}$^{\star}$& \textbf{0.84}           & \textbf{0.75}          \\ \hline
sqrt      &    41234     & \textcolor{red}{TO}              & \textcolor{red}{N.A.}$^{\star}$& \textbf{0.70}           & \textbf{0.53}          \\ \hline
square    & 35685        & \textcolor{red}{TO}              & \textcolor{red}{N.A.}$^{\star}$& \textbf{0.59}           & \textbf{0.55}           \\ \hline
arbiter    &  23619      & 59.53          & 65.56          & \textbf{0.38}           & \textbf{0.12}   \\ \hline
sin        &  8948      & \textcolor{red}{TO}              & \textcolor{red}{N.A.}$^{\star}$& \textbf{0.15}           & \textbf{0.13}           \\ \hline
adder      & 2548       & 4.25           & 53.4           & \textbf{0.04}           & \textbf{0.04}           \\ \hline

\textbf{GEOMEAN} & \textbf{-} & \textbf{-} & \textbf{-} & \textbf{0.65} & \textbf{0.46} \\ \hline

\end{tabular}}

\begin{tablenotes}
\item $^\star$ \footnotesize{Not available due to the prior e-graph to circuit conversion failure.}
\end{tablenotes}

\end{table}

\subsection{Efficient DAG-to-DAG Conversion}\label{exp:preprocess}

One bottleneck of the prior work \deftextEsyn is the conversion between circuit and e-graphs. Here we compare the circuit--e-graph mutual conversion time using the benchmark circuits.
As can be seen from \Cref{tab:time-data}, in most cases, our direct DAG-to-DAG conversion takes only a fraction of a second. Whereas, \deftextEsyn  hardly scales to over $10^4$ e-nodes, resulting in time-out in most cases.
The results indicate that our conversion method is minimally affected by the circuit size. Even when applied to circuits over $10^5$ e-nodes, the conversion can still be performed in a remarkably short runtime.
\vspace{-2pt}
\subsection{Enhancing Conventional Delay-oriented Synthesis }
\vspace{-2pt}
As delay-oriented optimization is crucial to meet timing constraints in logic synthesis,
in this experiment, we compare our approach against a competitive delay-oriented logic optimization flow published in \cite{sop}, which has been adopted by industrial users. The optimization sequence is as follows: \textit{\textbf{(st; if -g -K 6 -C 8)(st; dch; map)\textsuperscript{4}}}.
It takes advantage of SOP balancing~\cite{sop}, priority-cut-based mapping~\cite{mishchenko2007combinational} and resynthesis after mapping~\cite{structural_bias}.
We argue that it is nontrivial to outperform this flow with further optimization. 
For comparison, the \deftextEsynTurbo flow applies the same \textit{\textbf{(st; if -g -K 6 -C 8)(st; dch; map)\textsuperscript{3}}} first.
Before the final round of \textit{{\textbf{(st; dch; map)}}},
we make use of e-graph rewriting to obtain a set of structurally diverse solutions.
The results are shown in \Cref{tab:all-results}. Compared to the conventional delay-oriented optimization flow, \deftextEsynTurbo achieves delay reduction on all designs, averaging 7.29\%, along with an area saving of nearly 12.54\% while the runtime overhead is moderate. Due to space limit, we omit the result of  \deftextEsyn~\cite{chen2024syn} as it already times out for all large-scale circuits in this benchmark.
\subsection{GNN-based Cost Model for Faster Extraction}\label{sec:gnn-based-cost-model}

To further reduce the runtime overhead of \deftextEsynTurbo, one can plug in a machine learning model for rapid cost evaluation.
Here we demonstrate the use of state-of-the-art HOGA~\cite{deng2024hoga} model.
We firstly train it on circuits from the OpenABC-D benchmark~\cite{chowdhury2021openabcd} synthesized by the aforementioned  flow to adapt its prediction towards the setting in this experiment.
Each design module in the OpenABC-D benchmark generates 100 different structural samples, resulting in a training set of around 40,000 samples. 
The labels are obtained using the same ASAP 7nm cell library for technology mapping. The delay prediction achieves a Mean Absolute Percentage Error (MAPE) of 25.2\% and a Kendall's $\tau$ of 0.62. Using this prediction model, we present logic synthesis results in the last column of \Cref{tab:all-results}, which shows an averaged runtime saving of nearly 28\%.


\subsection{Runtime Analysis}

\begin{figure}[thb]
    \centering
    \begin{minipage}{0.49\linewidth}
        \centering
        \includegraphics[width=\textwidth]{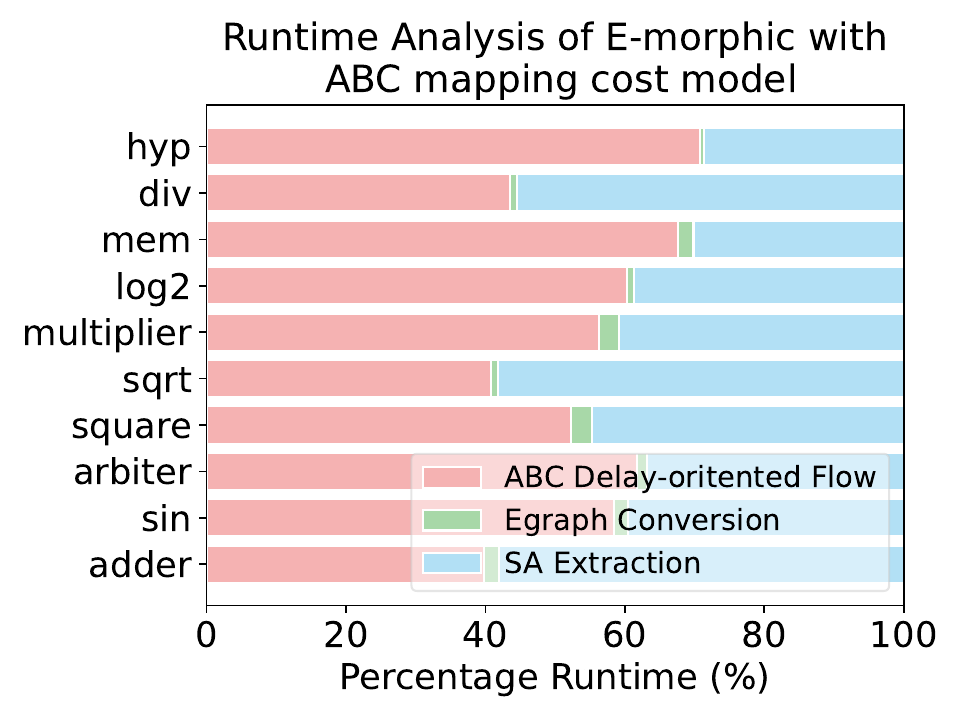}
    \end{minipage}
    \begin{minipage}{0.49\linewidth}
        \centering
        \includegraphics[width=\textwidth]{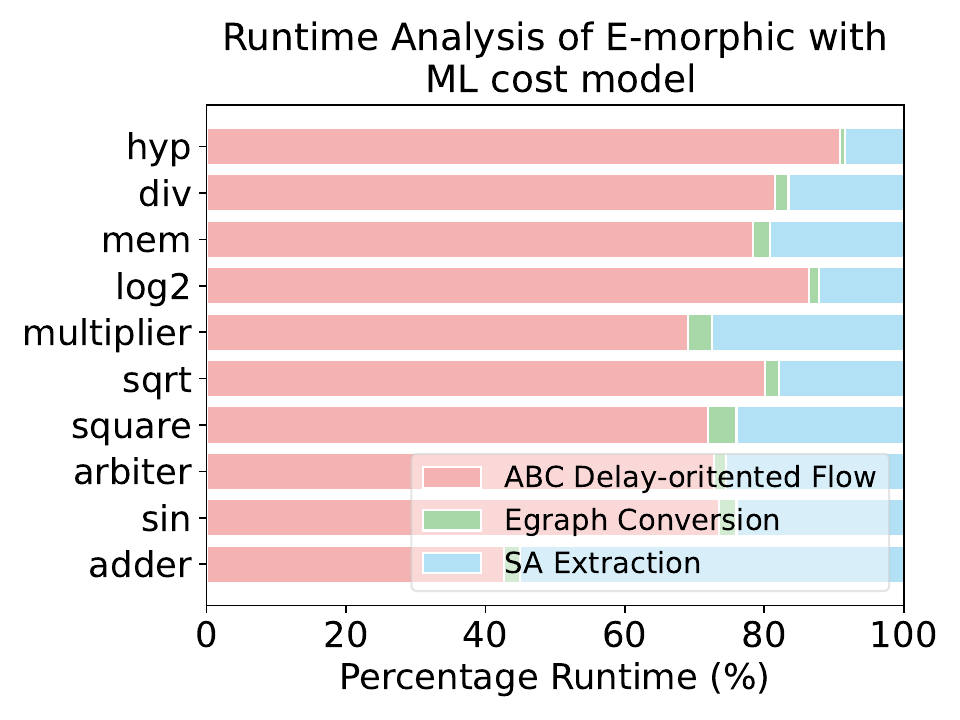}
    \end{minipage}
    \caption{Runtime breakdown of   \deftextEsynTurbo logic synthesis}
    \label{fig:runtime}
\end{figure}

We plot the proportion of time consumption of each step in \Cref{fig:runtime}.
It can be seen that the time spent on the existing technology-independent optimization and mapping in the delay-oriented flow~\cite{sop} accounts for a large portion of the total runtime, which is irrelevant to the usage of e-graph optimization. While the time for the additional operations in \deftextEsynTurbo (mainly conversion and simulated annealing)  is  moderate, demonstrating the efficiency of our framework. Particularly, this efficiency is more pronounced in large-scale circuits, where the proportion of our time overhead is significantly lower.

\section{Conclusion}
\label{sec:conclusion}

This paper proposes a scalable framework named \deftextEsynTurbo , which uses equality saturation for structure exploration before technology mapping. It features  efficient e-graph-circuit conversion, solution-space pruning and simulated annealing for extraction. Additionally, it allows the integration of ML prediction models  to further reduce the runtime.  The techniques presented in this paper will not only benefit applications of equality saturation in logic synthesis, but also other e-graph-based large-scale optimization problems.
\newpage
\bibliography{refs}
\balance

\end{document}